\documentclass[
showpacs,
nofootinbib,amsmath,amssymb]{revtex4}

\usepackage{graphicx}
\usepackage{dcolumn}
\usepackage{bm}
\usepackage{float}
\usepackage[hypertex]{hyperref}
\usepackage{hyperref}
\usepackage{amssymb}
\newcommand{\der}[1]{\frac{\partial}{\partial #1}}
\newcommand{\wtd}[1]{\widetilde #1}

\newcommand{\sss}[1]{\scriptscriptstyle #1}
\newcommand{\Ref}[1]{(\ref{#1})}

\newcommand{\be}{\begin{equation}}
\newcommand{\ee}{\end{equation}}

\begin{document}

\title{True self energy function and reducibility in effective scalar
        theories. \\ (Revised).}

\author{Vladimir V.~Vereshagin}
\email{vvv@AV2467.spb.edu}
\affiliation{St.-Petersburg State University, St.-Petersburg,
Petrodvoretz, 198504, Russia}



\begin{abstract}

This is the revised version of Sect. I - IV of the paper
\cite{VV2}
originally published in 2014. The thing is that in
\cite{VV2}
the text was insufficiently clear and, in addition, it contained (through my
fault) a few typos. This is the reason why I decided to offer a revised version.

\end{abstract}


\maketitle


\section{Introduction}
\label{Sec_introduction}

First of all it is necessary to recall the definition of the term
``effective theory'' suggested in
\cite{AVVV2}
and used throughout the paper.
{\it The theory is called effective if the corresponding interaction Lagrangian
in the interaction picture contains all the local monomials consistent with a
given linear symmetry}%
\footnote{This is just a slight modification of the definition
suggested in
\cite{WeinEFT};
see also the monograph
\cite{Weinberg1}.}.
In the paper
\cite{KSAVVV2}
it was given the definition of the
{\it effective scattering theory}:
this is just an effective theory only designed for calculating the
S-matrix (not Green functions). As pointed in
\cite{WeinAsySafe},
the Green functions, as well as effective Lagrangian, depend on the
infinite set of redundant parameters%
\footnote{This is just because an infinite set of different
Lagrangians  may result in the same S-matrix; see, e.g.
\cite{Tyutin},
\cite{equivalence_we}.}
(see, e.g.,
\cite{Weinberg1}),
while the S-matrix elements only depend on the essential parameters.
What is important, is that when the essential parameters are
concentrated in a certain area it looks possible to construct the
renormalizable S-matrix (see
\cite{WeinAsySafe}
and also
\cite{GomisWein}).
For this reason I find it interesting to make an attempt of
constructing the iteration scheme suitable for effective scattering
theory. Such a scheme should result in finite expressions for the
S-matrix elements at every step of the iteration procedure.
{\it The
finiteness of Green functions (off the mass shell) is not required;
it is only required the finiteness of their residues at}
$p_i^2 \to m_i^2$.

The obvious problem emerging immediately on this way is that of the two-point
Green function (self energy). In contrast to the S-matrix elements we need to
know this function off the mass shell. One more problem manifests itself when
one performs the conventional Dyson summation of the chain of two-point
functions to obtain the full propagator. The point is that the result
demonstrates the obvious contradiction with the limitation imposed by the
famous K\"{a}llen-Lehmann representation. Besides, when inserted in the
external line of a Green function of the multi-scalar effective scattering
theory, the 2-leg graph of brings unwanted poles which make the physical
interpretation contradictory. At last, the presence of many similar particles%
\footnote{Particles with identical quantum numbers except mass.}
in a theory makes the problem of diagonalization difficult.

The present paper is devoted to the discussion of above-mentioned problems
in the framework of the one-scalar effective scattering theory. My main goal is
to explain that when performing the renormalization it is much more convenient
to use the
{\em reduced}
graphs than to work with the graphs constructed from the initial Feynman rules.
I use (and explain when necessary) the terminology from the previous
publications (see
\cite{AVVV2},
\cite{KSAVVV2}
and references therein). I would like to stress that I work in terms of the
commonly accepted calculational scheme described in many text-books and
monographs (see, e.g.,
\cite{Weinberg1}, \cite{Bogoliubov}, \cite{Peskin}).

Three notes are in order. First, it is implied that in the theory considered
below there is no massless particle. This eliminates infrared problems.
Second, as usually the diverging integrals are considered regularized by
one-parametric cutoff. At last, I only consider the case of space-time
dimension
$D=4$.
Below I often use the following commonly accepted abbreviations:
1PI -- one-particle-irreducible, 1PR - one-particle-reducibile,
LSZ -- Lehman-Simanzik-Zimmerman, RP -- renormalization prescription.



\section{Preliminary notes}
\label{prelim}

First of all it is necessary to recall the reader some results
obtained in the previous papers (see
\cite{AVVV2}, \cite{KSAVVV2}, \cite{VV1})
and the terminology introduces therein. For simplicity I consider
here only the case of scalar theories. I refer the reader to the
above-cited papers for more detailed discussion and the relevant
figures.

In the papers
\cite{AVVV2}, \cite{KSAVVV2}
it has been considered the phenomenon of disappearance of the pole
associated with the propagator line of a particle with mass
$m$
and momentum
$p$
in S-matrix graph due to the presence of ``killing'' factors [those
proportional to
$(p^2-m^2)$]
in adjacent vertices and the corresponding confluence of these latter
ones. This phenomenon is called the
{\it reduction}
of a line. The vertex is called minimal with respect to its line
$p$
if it does not contain the corresponding
``killing''
factor
$(p^2-m^2)$.
The line
$p$
of a graph is called
{\it minimal}
if it cannot be reduced or, the same, if the adjacent vertex (or both
adjacent vertices if the line is inner) is minimal with respect to it.
The graph may be called minimal%
\footnote{This notion will be further refined when we consider the
4-leg graphs.}
if all its lines are minimal. Clearly, the reduction of all
internal lines of the a given graph results in the new graph that is
built entirely of
{\it minimal vertices}
each of which is minimal w.r.t. all its lines. Note that the analytic
expressions that correspond to the graphs under consideration (original and
reduced) are identical. I would like to stress that - by definition - all
subgraphs of the minimal graph are also minimal.

It can be easily understood that an arbitrary graph that provides the
nonzero contribution to
$S$-matrix
can be made minimal and, hence, it can only depend on the
{\it minimal}
parameters (coupling constants at the minimal vertices).
This means that the set of
{\it essential}
parameters only contains the minimal coupling constants. This set is
much more narrow as compared to the total number of coupling constants
(minimal plus non-minimal) of the effective theory. Nevertheless, it
is still infinite. This follows from the fact that all the vertices
of the form
$g_n \phi^n(x)$ ($n=5,6, \ldots$)
are minimal. The theories that contain vertices of these types
($n \geqslant 5$)
are nonrenormalizable. This means that one needs to attract an
infinite number of counterterms constructed from the field and its
derivatives of arbitrary order to eliminate the occurrence of
infinities in
$S$-matrix
elements. Hence, it is necessary to formulate an infinite number
of corresponding RP's including those fixing the finite parts of
non-minimal parameters. It turns out that the renormalization of the
$S$-matrix
graph constructed from the minimal vertices, in principle, might
introduce dependence on non-minimal parameters. This contradicts to
what is written above. Is there any way out of this contradiction? I
think the answer is yes. It is necessary to reconstruct the
renormalization procedure in such a way that the need in fixing the
non-minimal counterterms would not appear at all. Surely, this might
be only possible if the non-minimal coupling constants are certain
functions of the minimal ones. In other words, the renormalizability
of effective scattering theory requires the existence of certain
complicated symmetry that establishes linkage between the values of
different coupling constants. In this case  it looks like the number
of independent essential constants in the effective theory with a
single scalar particle should equal three: two minimal coupling
constants
$g_3$,  $g_4$
and the physical mass
$m$.
To chech/prove this guess it is necessary to construct the explicit
form of the corresponding symmetry relations
$$
F(m^2,g_3, g_4,\ldots) = 0.
$$
Here I imply that the set of arguments of the function
$F$
contains all parameters that appear in the basic Lagrangian (both minimal and
nonminimal).

In this paper I make the very first step on the way of constructing the
relevant renormalization procedure. I follow the conventional logical scheme.
First of all one needs to perform the renormalization of 1PI one-loop
$n$-leg
graphs for
$n=1,2,,3,\ldots$.
Then these renormalized (finite!) graphs can be considered as subgraphs in the
structure of 1PI 2-loop n-leg graphs which, in turn, must be renormalized, and
so on. The new feature that manifests itself in the case of effective
scattering theory is the emergence of possibility to introduce two different
definitions of one-particle irreducibility -- the graphical or the analytical
1PI. This problem is discussed in the
Sec.~\ref{OP_2leg}


\section{The most general form of local vertices}
\label{OPvertices}

Let us first consider the simplest effective theory: that containing only one
real scalar field
$\phi(x)$:
\be
\phi(x) = \frac{1}{(2\pi)^3}\int \frac{d^3p}{2p_{\,0}}\:
[a^+(p)\exp({ip\,x}) + H.c.].
\nonumber
\ee
The creation and annihilation operators fulfil the conventional commutation
relation
\be
[a^-(p),a^+(q)]_- =
(2\pi)^3\,2p_{\sss 0}\; \delta ({\bf p}- {\bf q})\, .
\nonumber
\ee
Here
$
p_{\,0} = \sqrt{{\bf p}^{\,2} + m_k^2}
$
and
$m$
stands for the physical mass.

Note that I rely upon the renormalized perturbation scheme with OMS
(on-mass-shell) renormalization prescriptions. R-operation is precisely that
described in
\cite{Collins}
(see also
\cite{vasiliev2}).

The full interaction Lagrangian density of the effective theory is the sum of
an infinite number of local terms of the form
\be
L_{int}(x) = \sum_{n=0}^\infty \left[L_n(x) + C_n\right]\, ,
\label{hamiltonian}
\ee
where
$L_n(x)$
is an infinite sum of
{\it all}
Lorentz-invariant
$n$-leg
local vertices constructed from the field and its derivatives of various
orders.
$C_n$
stands for the full sum of $n$-leg counterterms.

To present
$H_n$
in explicit form it is necessary to introduce a contracted notation for the
field derivatives of various orders. Let us define
$$
\partial^{[s]} \stackrel{\rm def}{=}
\partial^{\mu_1}\! \ldots\, \partial^{\mu_s}.
$$
The most general triple interaction Lagrangian density may be written as an
infinite sum of local terms of the form
\be
L_3 = \frac{1}{3!} \sum_{s=0} {\tilde D}^{jk;s}
    :\phi
    \left(\partial^{[s]}\phi^j\right)
    \left(\partial_{[s]}\phi^k\right):\; ,
\label{triple_OPvertex}
\ee
where
$: \ldots :$
denotes the normal product,
\begin{gather}
\phi^i \stackrel{\rm def}{=} K^i \phi\; ,  \nonumber \\
K \stackrel{\rm def}{=}
-(\partial^{\mu}\partial_{\mu} + m^2)\; , \nonumber \\
K^i \stackrel{\rm def}{=}
\underbrace{K \ldots K}_{i\; \rm times}\, , \nonumber
\end{gather}
and
$
{\tilde D}^{jk;s}
$
are real (dimensional) coupling constants. In
\Ref{triple_OPvertex}
there are no derivatives acting on the field
$
\phi
$
because one can make use of the integration by parts.

For the following we do not need to know the form of vertices with
$
l > 3
$
lines. Nevertheless it may be useful to show how one can write down, say, the
vertex with four lines:
$$
L_4 = \frac{1}{4!}
\sum_{ijk}^{\infty} \sum_{s_1s_2s_3}^{\infty}
           {\tilde D}^{\; ijk;s_1s_2s_3}
 :\phi \left(\partial^{[s_1]}\partial^{[s_2]}\phi^i \right)
         \left( \partial_{[s_2]}\partial^{[s_3]}\phi^j\right)
         \left( \partial_{[s_3]}\partial_{[s_1]}\phi^k\right):\; .
$$
The generalization for the case of
$
l>4
$
lines is straightforward.

In momentum space the Feynman rules needed to write down the 2-leg graphs are
constructed from the elements of bare propagator
$\pi$:
$$
\pi(p_i) = \frac{1}{p_i^2-m^2} ,
$$
and the vertices of the form :
\be
V(p_1, p_2, p_3) = i (2\pi)^4
\delta(p_1+p_2+p_3)\sum_{i,j,k=0}^{\infty}
D^{ijk}(p_1^2-m^2)^i (p_2^2-m^2)^j (p_3^2-m^2)^k \,
\label{OPtripleV}
\ee
(all the lines are considered internal). If, say, the line
$p_i, (i=1,2,3)$
is external, then
$p_i^2=m^2$
(recall that we only need to consider the one-loop 2-leg graphs!). Here
$
D^{ijk}
$
are just certain sums constructed from the above-introduced coupling constants
$
{\tilde D}^{jk;s}
$
and masses.

The 4-leg effective vertex of the Lagrangian level depends (symmetrically) on
three dependent Mandelstam variables (as above, when the line
$p_i$
is external,
$p_i^2=m^2$)
$s=(p_1+p_2)^2$, $t=(p_2+p_3)^2$, $u=(p_3+p_1)^2$;
$s+t+u=\sum_{i=0}^4 p_i^2.$
Of course, this is a manifestation of the 3-variable symmetry in
$\{s,t,u\}$.
In turn, this latter symmetry is associated with the original Bose symmetry
with respect to
$\{k_1,k_2,k_3\}$
and follows from the requirement of Lorentz symmetry.

To make formulae shorter I often use the notation
$$\kappa(k) \equiv (k^2-m^2)\, .$$

\section{The one-loop 2-leg function, self energy and irreducibility.}
\label{OP_2leg}

Using the above-given form
\Ref{OPtripleV}
one can construct the most general expression for the one-loop two-leg
function that is conventionally called as one-loop self energy. It reads
($k$
and
$q$
stand for incoming and outgoing momenta, respectively)%
\footnote{For the following discussion factors
$\frac{i}{(2\pi)^4}$
and common delta function are not essential and therefore omitted.}
\be
S(\kappa)= \sum_{ijklmn=0}^\infty D^{ijk}D^{lmn}\kappa^{i+l}
\int\! dr\,
(r^2-m^2)^{j+m-1} [(k+r)^2-m^2]^{k+n-1}
+ C(\kappa, \Lambda)\; .
\label{OPself_energy}
\ee
Here
$C(\kappa, \Lambda)$
stands for the counterterm series:
\be
C(\kappa, \Lambda) = \left[
C^{\rm [log]}(\kappa) \cdot {\rm log}\Lambda +
\sum_{n=0}^\infty C^{[n]}(\kappa) \Lambda^{2n}
\right] ,
\label{OP2_counterterms}
\ee
where
$\Lambda$
is the cutoff parameter  and every
$C^{[x]}(\kappa)$
$(x={\rm log},0,1,\ldots)$
is a power series%
\footnote{Note that the form
\Ref{OPself_energy}
is needed solely for subsequent using the LSZ formula that implies
$\kappa \neq 0$.
As shown below, this is not needed in the case of effective scattering
theory that is based on the Bogoliubov's scheme.}
in
$\kappa$:
\be
C^{[x]}(\kappa) = \sum_{n=0}^\infty c_n^{[x]} {\kappa}^n .
\label{couterterm_structure}
\ee
Recall that in effective theory
\underline{all}
the types of two-leg counterterms are presented in
\Ref{OP2_counterterms}.
The counterterms of the types
$C^{[x]}(\kappa)$
$(x={\rm log}, 1,2,\ldots)$
are needed to remove infinities, while
$C^{[0]}(\kappa)$
are used for the finite renormalization  required by RP's.

It can be easily shown that the sum
\Ref{OPself_energy}
contains only one nontrivial integral (it corresponds to
$j+m=k+n=0$):
\be
I^{0,0}(\kappa) = \int\!  \frac{dr}{(r^2-m^2) [(k+r)^2-m^2]}\,
 \equiv
\left[ J(\kappa) + a_1 {\rm log} \Lambda + a_2 \right]\, .
\label{OPmain_int}
\ee
(recall that the common delta-function is omitted). All the other integrals
diverge like the powers of
$\Lambda$.
In
\Ref{OPmain_int}
$a_1$ and $a_2$
are just arbitrary constants (depending on
$m^2$)
while the integral
$J(\kappa)$
is understood as the part of
$I^{0,0}(\kappa)$,
which remains after the infinite renormalization is done. Of course, this part
depends on all finite counterterms.

Clearly, infinite counterterms in
\Ref{OP2_counterterms}
can be chosen so that they cancel all the infinite
contributions. The finite parts should be chosen in accordance with the
normalization conditions. So, the expression
\Ref{OPself_energy}
can be rewritten as follows:
\be
S(\kappa)= \sum_{il=0}^\infty D^{i00}D^{l00}
\kappa^{i+l} J(\kappa) + \sum_{n=0}^\infty c_n {(\kappa)}^n\, .
\label{OP_SE_fin}
\ee
Here
$c_n$
are the new (finite) counterterm coefficients to be fixed with the help of
renormalization prescriptions. Let us present
\Ref{OP_SE_fin}
in the form most suitable for the following analysis. For this it is convenient
to reorder the terms in
\Ref{OP_SE_fin}
as follows:
\be
S(\kappa)= \sum_{i=0}^\infty G^i J(\kappa) {\kappa}^i +
\sum_{i=0}^\infty \wtd{d_i}{\kappa}^i\, .
\label{OPfinal_SE_fin}
\ee
Here
\be
G^i = G^0 + 2\sum_{k=1}^i D^{k00}D^{(i-k)00}\, ,
\label{OP_D-G}
\ee
and the coefficients
$\wtd{d_i}$
(free parameters!) are certain combinations of
$c_n$
and various degrees of
$m^2$.

The problem is that the number of unknown parameters
$\wtd{d_i}$ $(i=0,1,2,\ldots)$
in our theory is actually infinite, while we have only two physically
motivated restrictions that can be used to fix them. They are the
following:
\be
\left. S(\kappa)\right|_{\kappa=0} = 0\,
\label{OP_RP_mass}
\ee
(fixes the pole position of the 2-leg Green function), and
\be
\left.\der{\kappa}S(\kappa)\right|_{\kappa=0} = 0\,
\label{OP_field_norm}
\ee
(fixes the residue at pole or, the same, normalization of the wave function;
the normalization corresponding to
\Ref{OP_field_norm}
is especially convenient for dealing with S matrix). Let us try to fulfil
formally these restrictions and analyze the results. Substituting
\Ref{OPfinal_SE_fin}
in
\Ref{OP_RP_mass}
we obtain:
\be
\wtd{d_0} = - G^0 J(0)\, .
\label{OP_massCT}
\ee
Then, from
\Ref{OP_field_norm}
it follows:
\be
\wtd{d_1} = G^0 J'(0) - G^1 J(0)\, .
\label{OP_fieldCT}
\ee
So, the counterterm coefficients
$\wtd{d_i}$
with
$i \geqslant 2$
remain unfixed (recall that they are certainly nonzero).

Here is a point to remind the reader that both the requirements
\Ref{OP_RP_mass}
and
\Ref{OP_field_norm}
are based on the result of formal computation of the exact (or. the same, full)
propagator
$P(p^2)$
by way of summing Dyson's chain constructed from an infinite number of links
(2-leg insertions) connected with one another by the simple propagator%
\footnote{I would like to stress that at this point it is tacitly
assumed that every interim propagator is really presented it the
chain. This is not always the case in effective theory just because
some of them might be "killed" by the corresponding factors stemming
from the adjacent vertices. For this reason it turns out possible to
rely upon the alternative definition for the notion 1PI.}.
Every link is considered as the 1PI full 2-leg function
$S$
(conventionally called ``self energy''):
\be
P(p^2) = \pi + \pi S\, \pi + \pi S\, \pi S\, \pi + \ldots
 \stackrel{\rm formally}{=}
\frac{\pi}{1-\pi S} =
\frac{1}{\kappa - S}\, .
\label{PO_full_prop}
\ee
The result in the RHS of
\Ref{PO_full_prop}
is only valid under the condition that%
\footnote{The violation of this condition was a key point that
allowed Veltman (see
\cite{Veltman})
to obtain his famous conclusions concerning the description of
unstable particles in the framework of QFT.}
\be
\left|\frac{S(\kappa)}{\kappa} \right| < 1.
\label{OP_sigma growth}
\ee
In familiar renormalizable theories this limitation is certainly
fulfilled. That is why in such case the conditions
\Ref{OP_RP_mass}
and
\Ref{OP_field_norm}
can be used as legitimate RP's. However, this is not true in the case
of effective theory. To show this let us make use of the requirement
that follows%
\footnote{In fact, this is just a version of the well known
consequence of K\"{a}llen-Lehmann representation (see  Chapter 10.7
in the monograph
\cite{Weinberg1}).}
from
\Ref{OP_sigma growth}:
\be
{\Bigl|} S(\kappa) {\Bigr|}_{\kappa \rightarrow\infty}< \kappa\, .
\label{OP_growth1}
\ee
If this limitation is broken the use of RP's
\Ref{OP_RP_mass}
and
\Ref{OP_field_norm}
as the normalizing conditions for 2-leg function turns out groundless.

There is a different argumentation (not based on the full summing of
Dyson's chain) in favor of using those RP's for the normalization of
2-leg function. It is based on the quite natural requirement: neither
the pole location nor the residue should be changed by the higher
orders of the loop expansion. This argumentation is no less correct
than that discussed above. The problem is that in effective theory
the straightforward using of RP's
\Ref{OP_RP_mass}
and
\Ref{OP_field_norm}
looks a bit naive since it certainly leads to unsatisfactory result.

Note that the expression
\Ref{OP_fieldCT}
requires attracting the RP for the non-minimal parameter
\be
G^1 = 2D^{000}D^{100}.
\label{G1}
\ee
It can be shown that the renormalization of 3-leg one-loop graphs
would, in turn, require fixing the parameters
$G^i$
with
$i=2,3,\ldots$.
This contradicts to what has been written in
\cite{AVVV2}
(and compactly recalled in
Sec.~\ref{prelim}).
Similarly, as have been shown above, the direct summing of Dyson's
chain leads to the contradiction with K\"{a}llen-Lehmann
representation.

I think the reason for these problems lies in erroneous (naive) identification
of the expression for the one-loop propagator (2-leg Green function) following
from the effective theory Lagrangian with that for the one-loop 2-leg function
$\Sigma(p^2)$
which occurs in S-matrix graphs with arbitrary number of legs
$n \geqslant 1$
and
$l \geqslant 1$
loops. Such an identification seems me too forthright. In fact, we have to deal
with
\underline{two}
functions -
$G_2(k^2)$
and
$\Sigma(k^2)$.
These functions differ from each other:
$G_2(\kappa)$
is normalized near
$\kappa = 0$
by two conditions
\Ref{OP_RP_mass}
and
\Ref{OP_field_norm}
while
$\Sigma(\kappa)$ -
by the only condition
\Ref{OP_RP_mass}
for arbitrary value (in the physical area)
$\kappa \geqslant 3m^2$.
Besides,
$\Sigma(\kappa)$
must satisfy the condition analogous to
\Ref{OP_growth1}
for the sake of considering Dyson's chains with arbitrary number of links. Of
course, it is implied that the infinite renormalization is done in both cases
and both functions may only depend on relevant finite counterterms.

The written above can be presented in the form of two equalities:
\be
G_2(\kappa) = J(\kappa) + g_0 + g_1\kappa + \sum_{n=2}^\infty
g_i\kappa^i,
\label{G2}
\ee
and
\be
\Sigma(\kappa) = J(\kappa) + \sigma_0 + \sigma_1\kappa +
\sum_{n=2}^\infty\sigma_i \kappa^i.
\label{Sigma}
\ee
Here
$g_i\,\, {\rm and}\,\, \sigma_i,\, (i=0,1,2,...)$
are the finite counterterm coefficients while
$J(\kappa)$
has been defined in
\Ref{OPmain_int}.
If we take now
\be
g_0=0, \ \ \ g_1=-J'(0), \ \ \ g_i=0,\ \ (i=2,3,...)
\label{renormirovkaG2}
\ee
and
\be
\sigma_0=-J(0), \ \ \ \sigma_i=0,\ \ (i=1,2,...)
\label{renormirovkaSigma}
\ee
then we obtain the functions
$G_2$
and
$\Sigma$,
properly normalized near the point
$k^2=m^2$.
The function
$\Sigma$
will then grow in accordance with the condition
\Ref{OP_sigma growth}.
This means that we can use that function which we find necessary (and
sufficient!) for the sake of computing the S-matrix graphs. It remains to prove
that such a function is
$\Sigma(\kappa)$.

To prove this, we must show that the expression
\Ref{Sigma}
(with relations
\Ref{renormirovkaSigma}
taken into account)
describes
\underline{all}
2-leg subgraphs that may occur when we consider an arbitrary S-matrix graph%
\footnote{This
is quite sufficient to perform the renormalization procedure.}.

The proof is trivial. As written above (see
\cite{AVVV2}, \cite{KSAVVV2}
and
Sec.~\ref{prelim} )
{\em this is so just because all internal lines of any
\underline{reduced}
(arbitrarily complex) graph are minimal}. The external lines of such S-matrix
graph are minimal because we need to know the matrix elements on shell only.
So, Dyson's chains (both finite and infinite) inside arbitrary (reduced!)
S-matrix graph only may consist of the minimal 2-leg subgraphs
$\Sigma(\kappa)$.
This would not be so if we isolated the subgraphs in an unreduced graph. In
particular, it would be necessary to introduce counterterms to two-leg
subgraphs with different numbers of derivatives on external lines.

The renormalization procedure is constructed as the series of steps:
(renormalization of the one-loop 1-, 2-,..., n-leg,... graphs). The analytic
expressions for the initial and reduced graphs are identical. So, Weinberg's
theorem on the high energy behavior
\cite{Weinberg1960}
is applicable for both types of one-loop graphs and one has to fix only minimal
counterterms. All this means that we can consider the 1-loop minimal 2-leg
graph
$\Sigma(\kappa)$
as the true
\underline{one-loop self energy function}.
Surely, the true
\underline{one-loop Green function}
defined by the relations (only near the pole location!) is G2.

\begin{figure}[t]
\includegraphics[height=2.5cm]{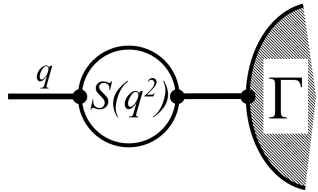}
\caption{One loop insertion in external line of the S-matrix graph. The
RP's for
$\Sigma(\kappa)$
should be taken according to
\Ref{Sigma}
and
\Ref{renormirovkaSigma}.}
\label{fig1}
\end{figure}

At this point it seems me useful to formulate compactly the sequence of steps
needed to perform the 1-loop renormalization of a given S-matrix graph. It
looks as follows.
\begin{enumerate}
\item
Draw the S-matrix graph
$G$
under consideration. This should be done in accordance with Feynman rules that
correspond to the
{\em initial}
effective Lagrangian. Select its 1PI one-loop subgraphs. (It is these subgraphs
that are necessary and sufficient for renormalization).
\item
Perform the reduction of all internal lines of the 2-leg subgraph
$S(r^2)$
selected at the previous step. This results in the sum of subgraphs of the same
loop order
$l=1$
as the initial ones. The loop numbers of these new subgraphs (we need to
preserve the initial loop counting rules!) should be computed as the sums of
the number of explicitly drawn loops plus the loop index (equal to 1) of the
formally pointlike secondary vertex constructed from the coupling constants of
the completely reduced subgraph
$\Sigma(r^2)$
(so-called
{\em secondary vertex}
of the one-loop level (see
\cite{AVVV2})).
\item
Add the relevant one-loop counterterms. Keep in mind that the secondary
vertices of one-loop level look precisely like the one-loop counterterms. Unite
both series. The result is nothing but a new (redefined) counterterm series.
\item
Impose the relevant (minimal!) RP's -- the result will be the correctly
renormalized one-loop two-leg graph.
\end{enumerate}

The advance of the above-described approach is obvious: one has no need in
formulating the non-minimal RP's for the one-loop 2-leg subgraphs!

\begin{figure}[t]
\includegraphics[height=7cm]{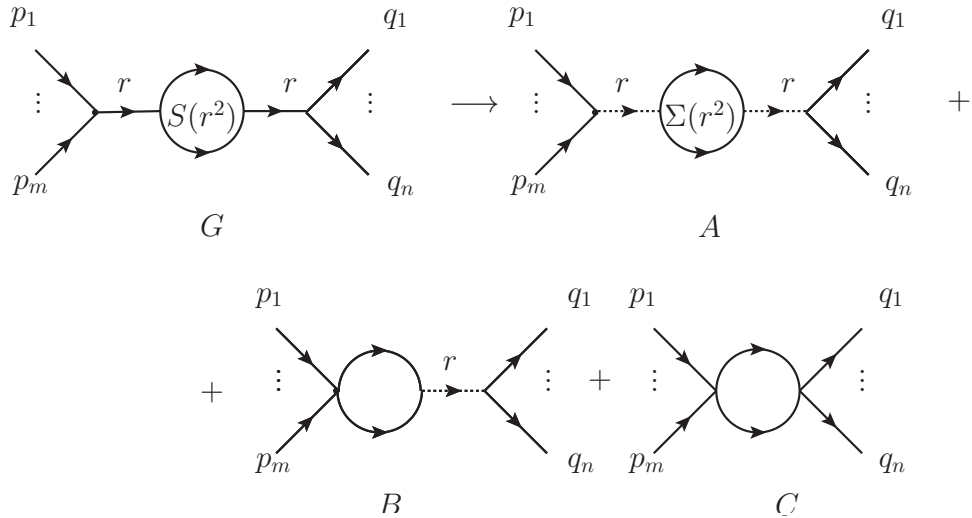}
\caption{The reduction of the lines
$r$
of the graph
$G$
that follows from the Feynman rules based on the original effective Lagrangian.
I do not show the relevant numerical coefficients in front of graphs
$A$, $B$
and
$C$
because they have no relation to the discussion of the topological structure.
The dotted lines
$r$
are minimal because the adjacent vertices are implied minimal with respect to
these lines. The solid lines correspond to the sums of contributions of both
kinds -- minimal and nonminimal. }
\label{fig2}
\end{figure}

Clearly, the same logic applies to the case when we consider the effect caused
by the insertion of the two-leg one loop graph
$S(r^2)$
in internal line (see the graph
$G$
in the left side of
Fig.~\ref{fig2}).
It is clearly visible that the reduction of both lines
$r$
(this does not change the analytical expression of the graph
$G$!)
leads to a separation of the initial graph into three parts. Only one of these
parts requires knowledge of the one loop 2-leg graph
$S(r^2)$;
two others relate to the next steps of the renormalization (3-, 4, ...,
n-leg graphs). This part is presented by the graph
$A$,
where the 2-leg subgraph is precisely
$\Sigma(r^2)$.
The graphs
$B$
and
$C$
should be renormalized at next steps!

That is why (as have been explained above) one can introduce the alternative
(``analytical'') definition of what is irreducibility: graphical (G1PI)
versus analytical (A1PI) reducibility.

It might be useful to analyze the process of reduction of the simple
(undressed) internal line%
\footnote{Recall that there is no need in considering the reduction
of external lines because in
$S$-matrix
graphs all the external lines are minimal.}.
This point has been discussed already in papers
\cite{AVVV2}
and
\cite{KSAVVV2}
which I refer the reader to.

What is the essence of the above analysis? The thing is that for the
renormalization of a given graph it is necessary (and sufficient) to
renormalize all its 1PI subgraphs. In the case when we rely on the
G1PI concept we would need to fix the non-minimal counterterms (just
because the subgraphs may be non-minimal). In contrast, when all the
lines of the graph in question have been reduced (the graph is made
minimal), all its subgraphs turn out minimal and one only needs to
fix the minimal counterterms. This confirms the general logical line
described in
\cite{AVVV2}, \cite{KSAVVV2}.


\end{document}